\documentclass[aip,jap,preprint]{revtex4-1}

\usepackage{graphicx} 
\usepackage{amsmath}
\usepackage{epstopdf}

\usepackage{float}

\usepackage{color} 

\begin{document}

\title{Method for characterizing bulk recombination using photoinduced absorption}

\author{Nora M. Wilson}
\email[]{nwilson@abo.fi}
\author{Simon Sand\'en}
\author{Oskar J. Sandberg}
\author{Ronald \"{O}sterbacka}

\affiliation{Physics, Faculty of Science and Engineering and Centre for Functional Materials, \AA bo Akademi University, Porthansgatan 3, 20500 Turku, Finland}

\date{\today}

\begin{abstract}
The influence of reaction order and trap-assisted recombination on continuous-wave photoinduced absorption measurements is clarified through analytical calculations and numerical simulations. The results reveal the characteristic influence of different trap distributions and enable distinguishing between shallow exponential and Gaussian distributions as well as systems dominated by direct recombination by analyzing the temperature dependence of the in-phase and quadrature signals. The identifying features are the intensity dependence of the in-phase at high intensity, $\textit{PA}_\text{I}\propto I^{\gamma_\text{HI}}$, and the frequency dependence of the quadrature at low frequency, $\textit{PA}_\text{Q}\propto \omega^{\gamma_\text{LF}}$. For direct recombination $\gamma_\text{HI}$ and $\gamma_\text{LF}$ are temperature independent, for an exponential distribution they depend on the characteristic energy $E_\text{ch}$ as $\gamma_\text{HI}=1/(1+E_\text{ch}/kT)$ and $\gamma_\text{LF}=kT/E_\text{ch}$ while a Gaussian distribution shows $\gamma_\text{HI}$ and $\gamma_\text{LF}$ as functions of $I$ and $\omega$, respectively. 

\end{abstract}

\maketitle

\section{Introduction\label{introduction}}

Using disordered semiconductors, such as organic semiconductors, for building optoelectronic devices calls for a thorough understanding of recombination processes in these materials. For investigation of these processes in the bulk of a sample, continuous-wave photoinduced absorption (abbreviated cwPA or PIA) is a suitable choice as it is an optical experiment performed without contacts. It is a pump-probe technique achieving high sensitivity by modulating the excitation light at frequency $\omega$ and processing the resulting absorption change with a phase-sensitive lock-in amplifier. This does however also require the data to be analyzed in the frequency domain, making interpretation  challenging. A well-established use of cwPA,\cite{tauc-connor-1982, botta, dellepiane,wohlgenannt} alongside transient absorption measurements, \cite{nogueira-sm-2003,nogueira-jpcb-2003,montanari,orenstein} is examining long-lived excitations and their decay. Contrary to many experimental techniques used to study recombination, as for example light-intensity dependent open-circuit voltage measurements ($V_\text{oc}$) \cite{nyman,kuik,tress2013} and charge-extraction measurements, \cite{kirchartz2011,burke} photoinduced absorption is uninfluenced by processes at the contacts, such as band bending, injected carrier density profiles, and surface recombination. \cite{kirchartz2013,oskar2016,anton2016,deibel} As cwPA is performed under similar physical circumstances as $V_\text{oc}$-measurements the results are directly comparable. { The experimental setup of cwPA is almost identical to that of photoreflectance, where the photomodulated reflectance is measured to obtain the dielectric contant and draw conclusions about band-structure.\cite{yu-cardona,pollak} In contrast, we want to use cwPA, where the photomodulated absorption is measured, to obtain the carrier density and draw conclusions about the recombination.} \\

The reaction order, defined as the exponent of the charge carrier density in the recombination rate ($R\propto n^\delta$), is often used to describe recombination. For recombination between two free carriers, henceforth referred to as bimolecular recombination, the reaction order is $\delta=2$ while Auger type recombination gives $\delta=3$. 
Some materials do, however, show reaction orders $2<\delta<3$, for instance measurements on some polymer:PCBM blends indicate close to $\delta=2.5$, which has been attributed to the effectively two-dimensional morphology of the material\cite{juska2009,nyman} or trap-assisted recombination.\cite{shuttle} The influence of trap-assisted recombination is often taken into account through an exponential trap distribution. \cite{tachiya2002,tachiya2010,kirchartz2012}\\ 

In this work we clarified the impact of trap-assisted recombination on cwPA, with the aim of enabling investigation of bulk recombination using this contactless experiment. This was achieved through numerical simulations of cwPA based on an effective medium multiple trapping and retrapping model alongside analytical approximations. { The model utilizes the work by Westerling et al., \cite{westerlingI,westerlingII,westerlingIV,westerlingV} which focuses on using analytical approximation to study the impact of bimolecular recombination on cwPA. They also treat trap-assisted recombination in an exponential distribution, but mainly for transient behaviour. In our development of the method we expand the toolbox for analyzing trap-assisted recombination and include the case of an arbitrary recombination order.}\\

 The simulations encompassed three recombination mechanisms: direct recombination with the arbitrary reaction order $\delta$ as well as trap-assisted recombination in exponential and Gaussian trap distributions. In addition to distinguishing between these three mechanisms we also show how to extract the mean trap depth ($E_\text{ch}$) of an exponential distribution from the experimental data. { Using these results it is possible to identify the dominating bulk recombination mechanism both from intensity and frequency dependence measurements. For examples of how this method can be used see Sand\'{e}n et al.\cite{simon,simon2}}


\section{Model\label{model}}

In a photoinduced absorption experiment the change in transmission is measured under the influence of a pump light that generates charge carriers. In a thin film, for which the thickness $d$ and absorption coefficient $\alpha$ fulfill $\alpha d\ll1$, the relation between photogenerated charge carrier density $N$ and the relative change in transmission $\Delta\mathcal{T}/\mathcal{T}$ at a given probe wavelength is 
\begin{equation}
\frac{\Delta \mathcal{T}}{\mathcal{T}}=-\sigma d N,
\end{equation}
where $\sigma$ is the absorption cross section for the carriers at this wavelength.\cite{brabec} Replacing $d$ with $1/\alpha$ results in the corresponding equation for a thick film ($\alpha d\gg1$). Our simulations assumed that the probe wavelength was constant and that the change in transmission arose from a single type of excitation. \\

\begin{figure}[h]
 \includegraphics[width=8.5cm]{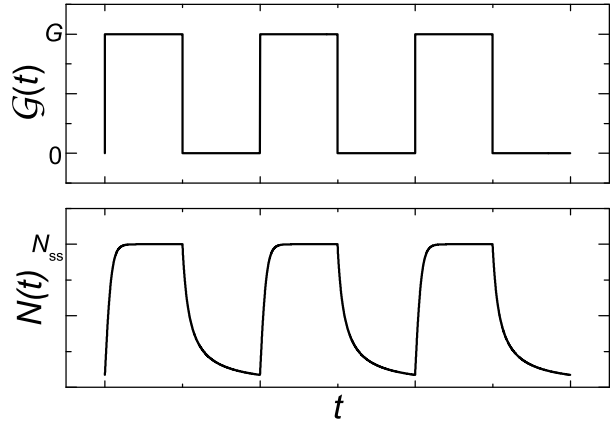}%
 \caption{\label{quasi-ss} Example of the generation and resulting charge carrier density. }
 \end{figure}

We simulated a cwPA experiment with the pump light intensity varying in the form of a square wave. Assuming linear generation (Intensity $\propto$ Generation), this leads to a generation rate of carriers $\mathcal{G}(t)$ that switches between 0 and $G$ with an angular frequency $\omega$. The resulting change in carrier density will become periodic, as illustrated in Figure~\ref{quasi-ss}. The transmission is analyzed as in-phase ($\textit{PA}_{\text{I}}$) and quadrature ($\textit{PA}_{\text{Q}}$), which have a frequency $\omega $ and a phase shift of 0$^\circ$ and 90$^\circ$ compared to the pump light, respectively. By expressing the components through a Fourier series they can, for generation with a sine wave as its first harmonic, be defined as \cite{westerlingV}

\begin{align}
\textit{PA}_\text{I}&=\frac{\omega\sigma d}{\pi}\int_c^{c+2\pi/ \omega}\!N(t)\sin{\omega t}~{\rm d}t \label{PAI-def}\\
\textit{PA}_\text{Q}&=-\frac{\omega\sigma d}{\pi}\int_c^{c+2\pi/ \omega}\!N(t)\cos{\omega t}~{\rm d}t,\label{PAQ-def}
\end{align}
where $c$ is an arbitrary constant. Equations~\ref{PAI-def}-\ref{PAQ-def} were used to simulate the cwPA results after calculating $N(t)$. \\

 \subsection{Direct recombination}

If an arbitrary reaction order $\delta$ describes the recombination and the thermal generation is negligibly small the density of carriers, $N(t)$, can be solved from 
\begin{equation}
\frac{{\rm d}N(t)}{{\rm d}t}=\mathcal{G}(t)-b N(t)^\delta,
\label{grundRO}
\end{equation}
where $b$ is a recombination constant. Note that $\delta$ can either be an actual reaction order, exactly describing the type of recombination mechanism, or an effective reaction order, giving an approximate description of underlying physics under certain physical circumstances. For direct recombination the recombination rate is $R=\beta n^2$, but a density dependent recombination constant $\beta(n)$ may result in $\delta\neq 2$. Equation~\ref{grundRO} formed the basis for our simulations of direct recombination, i.e. recombination between free charge carriers. \\

To analyze the impact of Equation~\ref{grundRO} on cwPA we used an effective lifetime $\tau_\delta$, corresponding to the lifetime of carriers at steady state under constant illumination. At steady state the density is $N_\text{ss}=(G/b)^{1/\delta}$ and rewriting the recombination term in Equation~\ref{grundRO} as $bN_\text{ss}^\delta=N_\text{ss}/\tau_\delta$ results in 

\begin{equation}\label{taudelta}
\tau_\delta\equiv b^{-1/\delta}G^{1/\delta-1}.
\end{equation}

Analytical expressions for the in-phase and quadrature with sinusoidal generation and $\delta=2$ have been presented by Westerling et al.\cite{westerlingII} We generalized these results to an arbitrary recombination order by replacing $\tau_\beta$ (describing bimolecular recombination) in their formulas with $\tau_\delta$, giving

\begin{equation}
\textit{PA}_\text{I}(G,\omega)= \left(\frac{\displaystyle G\tau_\delta}{\displaystyle 2}\right)\left(\frac{\displaystyle \sqrt{F(\omega,\tau_\delta)}}{\displaystyle (\omega\tau_\delta)^2+F(\omega,\tau_\delta)}\right)\sigma d\\
\vspace{-3mm}
\label{sinusoidalI}
\end{equation}\\
\begin{equation}
\textit{PA}_\text{Q}(G,\omega)=\left(\frac{\displaystyle G\tau_\delta}{\displaystyle 2}\right)\left(\frac{\displaystyle\omega\tau_\delta}{\displaystyle (\omega\tau_\delta)^2+F(\omega,\tau_\delta)}\right)\sigma d, 
\label{sinusoidalQ}
\end{equation}
where
\begin{equation}
F(\omega,\tau_\delta)={1-\frac{(\omega\tau_\delta)^2}{2}+\frac{1}{2}\sqrt{\Big((\omega\tau_\delta)^2+2\Big)^2-2}}.
\end{equation}
Looking at the limit $\omega\tau_\delta\ll 1$, i.e. near steady-state, {and noting that $\tau_\delta\propto G^{1/\delta-1}$} results in the in-phase intensity dependence 

\begin{equation}\label{HI-RO}
\textit{PA}_\text{I}\propto G^{1/\delta}.
\end{equation}

Westerling et al. conclude that {approximating the square-wave generation as sinusoidal in order to obtain} Equation~\ref{sinusoidalI} and \ref{sinusoidalQ} for the bimolecular case { introduces} an incorrect intensity dependence for the quadrature when $\omega\tau_\beta\ll1$.\cite{westerlingIV,westerlingV} Similarly, we concluded that Equation~\ref{sinusoidalQ} fails to describe the low frequency dependence for $\textit{PA}_\text{Q}$. For a better description of this behavior we have, as shown in the appendix, calculated an approximate solution for square wave generation given by 
\begin{equation}\label{LF-RO}
\textit{PA}_\text{Q}\propto\omega^{1/(\delta-1)}.
\end{equation}
Note that Equation~\ref{LF-RO} is inaccurate when $\delta$ is close to 2. When $\delta=2$ the behavior follows the relation $\textit{PA}_\text{Q}\propto\omega\ln{(\omega)}$, as shown by Westerling et al.\cite{westerlingV} \\

We define $G_0$ as the generation rate for which the lifetime $\tau_\delta(G=G_0)$ equals the inverse of $\omega$,
  \begin{equation}\label{G0}
G_0=(b^{1/\delta}/\omega)^{\delta/(1-\delta)}.
\end{equation} 
In a intensity (frequency) dependence measurement $G_0$ ($\tau_\delta$) will be a constant since $\omega$ ($G$) is fixed to $\omega_\text{fix}$ ($G_\text{fix}$). As $\omega\tau_\delta=1$ act as a divider between two regimes, differing by the lifetime being shorter or longer than a period of $\mathcal{G}(t)$, $\omega=1/\tau_\delta$ coincides with a notable change in the frequency dependence and $G_0$ with a change in the intensity dependence.\\

In addition to the results which can be explained by Equation~\ref{grundRO} experiments on disordered semiconductors normally display signs of dispersion.\cite{westerlingV,epshtein2001PRB} This dispersion is thought to originate from relaxation processes but a microscopic derivation of the implications on cwPA is lacking. It can be included through a distribution of lifetimes as suggested by Epshtein et al. or, as is often done in cwPA, phenomenologically through a Cole-Cole type function.\cite{epshtein2001PRB,colecole} We note that the in-phase is strongly affected by dispersion when  $\omega\tau_\delta\gg1$, thus making it a poor source of information of recombination mechanisms. Combined with the fact that the quadrature in this regime doesn't depend strongly on the reaction order, it is motivated to only focus on the high intensity and low frequency (near steady state) behavior.

\subsection{Trap-assisted recombination}

While Equation~\ref{grundRO} describes direct recombination, trap-assisted recombination doesn't follow a constant reaction order, instead it varies depending on the physical circumstances.  The rate for this recombination will depend on the density of free carriers ($n$ or $p$) and trapped carriers of opposite charge ($p_\text{t}$ or $n_\text{t}$) and be proportional to $pn_\text{t}$ or $np_\text{t}$.\\

\begin{figure}[h]
 \includegraphics[width=8.5cm]{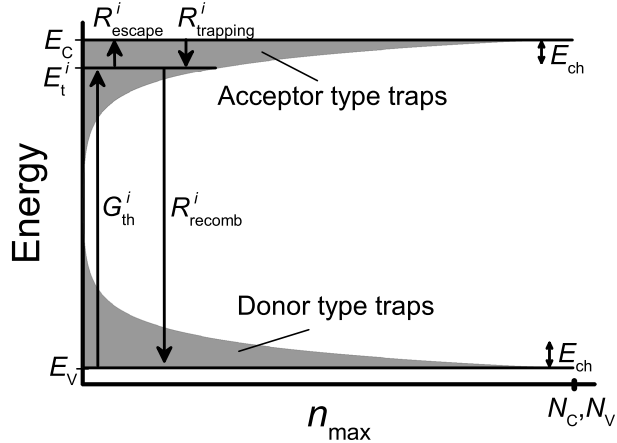}%
 \caption{\label{rates}Illustration of the energy levels and the rates in Equation~\ref{Eq-ntj}. The arrows indicate flow of electrons to and from traps at energy $E_\text{t}^i$.} 
 \end{figure}

The simulations of trap-assisted recombination used a multiple trapping and retrapping model utilizing Shockley-Read-Hall (SRH) statistics.\cite{shockleyread,hall1952} The energy levels comprise single level conduction and valence bands at $E_\text{C}$ and $E_\text{V}$ with trap distributions extending into the bandgap. Figure~\ref{rates} depicts this situation with an exponential distribution of traps. In the simulations we assume complete symmetry between electrons and holes. Therefore the distributions of acceptor and donor type traps, i.e. traps that are respectively negative or neutral when containing an electron, were symmetrical. 

In the simulations, four processes affected the density of trapped carriers: trapping, escape from traps, trap-assisted recombination, and thermal generation. Including these processes, in the mentioned order, to describe the density $n_\text{t}^i$ of electrons in acceptor traps at the energy $E_\text{t}^i$ gives \cite{wurfel,willemen}
\begin{align}\label{Eq-ntj}
\begin{split}
\frac{{\rm d}n_\text{t}^i}{{\rm d}t}=&\underbrace{\beta_1n(n_\text{max}^i-n_\text{t}^i)}_{R_\text{trapping}^i}-\underbrace{\beta_1N_\text{C}{\rm e}^{(E_\text{t}^i-E_\text{C})/kT}n_\text{t}^i}_{R_\text{escape}^i}\\
&-\underbrace{\beta_2nn_\text{t}^i}_{R_\text{recomb}^i}+\underbrace{\beta_2N_\text{V}{\rm e}^{(E_\text{V}-E_\text{t}^i)/kT}(n_\text{max}^i-n_\text{t}^i)}_{G_\text{th}^i}.
\end{split}
\end{align}
Here $n$ is the density of free electrons in the conduction band, $N_\text{C}$ ($N_\text{V}$) the density of states in the conduction (valence) band while $\beta_1$ and $\beta_2$ are capture rate coefficients. Note the use of symmetry which leads to $n=p$, $n^i_\text{t}=p^i_\text{t}$ and $n^i_\text{max}=p^i_\text{max}$. Figure~\ref{rates} illustrates the rates affecting the density. 

Describing the density of electrons in the conduction band ($n$) with the processes given in Equation \ref{Eq-ntj} results in 

\begin{align}\label{Eq-n}
\begin{split}
\frac{{\rm d}n}{{\rm d}t}=&~\frac{\mathcal{G}(t)}{2}+\sum_j^M[-R_\text{trapping}^j+R_\text{escape}^j-R_\text{recomb}^j+G_\text{th}^j]\\
&-\beta_\text{bim}( n^2-n_\text{i}^2),
\end{split}
\end{align}

\noindent where $\mathcal{G}(t)$ is the square wave generation (of charge carriers) switching between 0 and $G$ with the angular frequency $\omega$ and $M$ the number of trap levels. The constant $\beta_\text{bim}$ describes the bimolecular recombination and $n_\text{i}$ is the thermally induced intrinsic density of electrons in the valence band, assumed to be given by $n_\text{i}=N_\text{C}\sqrt{\text{e}^{-(E_\text{C}-E_\text{V})/kT}}$. The symmetry between donor and acceptor traps leads to every acceptor trap at $E_\text{t}^j$ corresponding to a donor trap at $E_\text{V}+E_\text{C}-E_\text{t}^j$ with the same value for $n_\text{max}^j$. After solving Equations~\ref{Eq-ntj}-\ref{Eq-n} numerically we calculated $\textit{PA}_\text{I}$ and $\textit{PA}_\text{Q}$ through Equations~\ref{PAI-def} and \ref{PAQ-def}. Since the intensity of the pump light and the frequency $\omega$ can be adjusted in the experimental setup these were also varied in the simulations. As the intensity is directly proportional to the generation $G$, results where $G$ is varied are referred to as intensity dependence. \\


Using different relations between $n^j_\text{max}$ and $E_\text{t}^j$ gives the opportunity to study different trap distributions. An exponential distribution of (acceptor type) traps is described by

\begin{equation}\label{ntexp}
n_\text{max}(E)=\frac{N_\text{t}}{E_\text{ch}}\exp{\left(\frac{E-E_\text{C}}{E_\text{ch}}\right)}.
\end{equation}
Here $N_\text{t}$ is the total trap density and $E_\text{ch}$ is the characteristic energy giving the mean trap depth. The corresponding distribution of donor traps is obtained by replacing $E$ with $E_\text{V}+E_\text{C}-E$. To use Equation~\ref{ntexp} in Equations~\ref{Eq-ntj}-\ref{Eq-n} we approximated the distribution with $100$ discrete trap levels, evenly spaced throughout the band gap, as this amount was found sufficient.


In addition to an exponential distribution we studied a Gaussian trap distribution, the trap density of which is 
\begin{equation}
n_\text{max}(E)=\frac{N_\text{t}}{\sigma_\text{G}\sqrt{2\pi}}\exp{\left(-\frac{(E-E_0)^2}{2\sigma_\text{G}^2}\right)},
\end{equation}
where $N_\text{t}$ is the total number of traps, $\sigma_\text{G}$ is the width of the distribution and $E_0$ is the energy at which the distribution peaks.

\section{Results\label{results}}
Through numerical simulations we clarified the intensity and frequency dependence for direct recombination with different reaction orders as well as for trap-assisted recombination in two trap distributions: exponential and Gaussian. { Note that all the plots in Figures~\ref{RO}--\ref{Gauss} are on a log-log-scale. Thus we can define a high intensity slope of $\gamma_\text{HI}$ for the in-phase by $\textit{PA}_\text{I}\propto G^{\gamma_\text{HI}}$ and a low frequency slope of $\gamma_\text{LF}$ for the quadrature by $\textit{PA}_\text{I}\propto \omega^{\gamma_\text{LF}}$.} Our simulations for shallow (40 meV) single-level traps showed no influence on the slopes and will therefore not be presented here. The parameters used in the simulations are listed in Table \ref{parameters}.

\begin{table}[H]
  \centering
{
  \caption{Parameters used in the simulations.}
    \begin{tabular}{llr}
    \hline
  Total density of traps &  $N_\text{t}$ & $10^{20}\text{ cm}^{-3}$ \\
   Density of states in conduction band &  $N_\text{C}$ & $10^{20} \text{ cm}^{-3}$ \\
    Density of states in valence band & $N_\text{V}$ & $10^{20} \text{ cm}^{-3}$ \\
  Absorption cross section for charge carriers &  $\sigma$ & $10^{-16} \text{ cm}^{2}$ \\
 Film thickness &   $d$   & $300 \text{ nm}$ \\
Value for $\omega$ in intensity dependence &    $\omega_\text{fix}$ & 628 $\text{s}^{-1}$ \\
 Value for $G$ in frequency dependence &   $G_\text{fix}$ & $2\times10^{21}\text{ cm}^{-3}\text{s}^{-1}$ \\
Constant in Equation~\ref{Eq-ntj} &    $\beta_1$ & $10^{-10}\text{ cm}^{3}\text{s}^{-1}$ \\
Constant in Equation~\ref{Eq-ntj} &    $\beta_2$ & $2.5\times10^{-12}\text{ cm}^{3}\text{s}^{-1}$ \\
Bimolecular recombination constant &    $\beta_\text{bim}$ & $5\times10^{-12}\text{ cm}^{3}\text{s}^{-1}$ \\
 Recombination constant in Equation~\ref{grundRO} &   $b$ & $(10^{-5}/G_\text{fix}^{-1+1/\delta})^{-\delta}$ \\
 Band gap &  $E_\text{g}$ & 1.2 \text{ eV} \\
 Maximum number of periods of $\mathcal{G}(t)$ simulated &   $N_\text{max}$ & 1000 \\
 Number of data points per period &   $m$     & $10^{5}$ \\
    \hline
    \end{tabular}%
  \label{parameters}%
}
\end{table}%

 \subsection{Direct recombination} 
\begin{figure}[h]
\includegraphics[width=8.5cm]{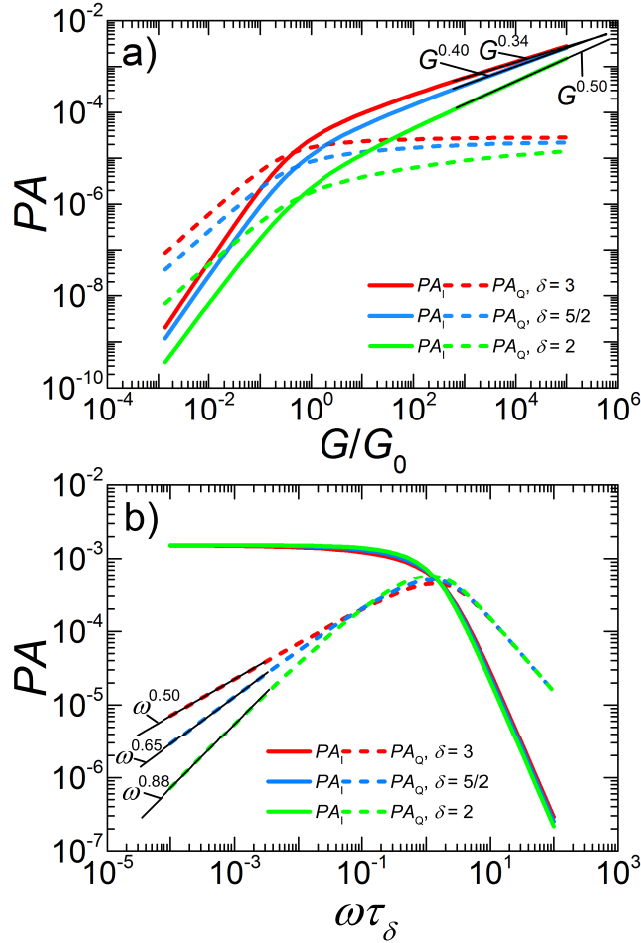}
 \caption{\label{RO}In-phase ($\textit{PA}_\text{I}$) and quadrature ($\textit{PA}_\text{Q}$) as functions of the generation $G$ (a) and frequency $\omega$ (b). Simulated results plotted for three reaction orders $\delta$. 
}
 \end{figure}

 Figure~\ref{RO} illustrates the intensity and frequency dependence for the three reaction orders $\delta=2$, 2.5 and 3, simulated using Equation~\ref{grundRO}. 
We use $\tau_\delta$ and $G_0$ to normalize $G$ and $\omega$, enabling better comparison of the plots. 

Figure~\ref{RO}(a) demonstrates that the high intensity slopes of the in-phase, {$\gamma_\text{HI}=0.50$} ($\delta$=2), 0.40  ($\delta$=2.5), and 0.34  ($\delta$=3), nicely agree with the predictions of Equation~\ref{HI-RO}, $\textit{PA}_\text{I}\propto G^{1/\delta}$. Similarily Figure~\ref{RO}(b) shows that the low frequency slopes { of the quadrature, $\gamma_\text{LF}=0.65$ ($\delta$=2.5) and 0.50  ($\delta$=3),} are in good agreement with Equation~\ref{LF-RO} for reaction orders $\delta>2$. The slope of 0.88 obtained for $\delta=2$ disagrees with Equation~\ref{LF-RO} as a consequence of the logarithmic behavior $\textit{PA}_\text{Q}\propto\omega\ln{\omega}$. The low intensity behavior in Figure~\ref{RO}(a) and the high frequency behavior in \ref{RO}(b) reveals that the present model, as expected, fails to reproduce the impact of dispersion, which according to the model by Epshtein et al. would result in the same slope for both the in-phase and the quadrature in this regime, where $\omega\tau_\delta\gg1$.\cite{epshtein2001PRB}

\subsection{Exponential trap distribution}
To interpret the results of an exponential trap distribution we used the analytical approximation presented by Westerling et al. \cite{westerlingI} which leads to the expression 
\begin{equation}\label{qnr}
n=N_\text{C}\left(\frac{n_\text{t}}{N_\text{t}}\right)^{E_\text{ch}/kT}.
\end{equation}
The derivation of this expression requires that the trapped carriers are in equilibrium with the free carriers and that $kT/E_\text{ch}\ll1$. Using Equation~\ref{qnr} for the relation between $n$ and $n_\text{t}$, trap-assisted recombination can be included through
\begin{equation}\label{Cnnt}
\frac{{\rm d}N(t)}{{\rm d}t}=\mathcal{G}(t)-2Cn(t)n_\text{t}(t),
\end{equation}
where the constant $C=\beta_2$ if $\beta_1\ll\beta_2$ (immediate trapping). If the density of traps is very large, $N\approx 2n_\text{t}\gg 2n$ and Equation~\ref{Cnnt} takes the same form as Equation~\ref{grundRO} with
\begin{align}
&\delta=1+\frac{E_\text{ch}}{kT} \label{RO-exp}\\
&b={C}N_\text{C}/(2N_\text{t})^{E_\text{ch}/kT}\label{beta-exp}.
\end{align}

The effective reaction order in Equation~\ref{RO-exp} is the same as that calculated by Kirchartz {and Nelson}.\cite{kirchartz2012} We have presented modelling of cwPA-results using Equation~\ref{qnr} and \ref{Cnnt} in another publication.\cite{simon} Combining Equation~\ref{RO-exp} with Equation~\ref{HI-RO} and \ref{LF-RO} results in 
\begin{align}
&\textit{PA}_\text{I}\propto G^{1/(1+E_\text{ch}/kT)}\label{HI-exp}\\
&\textit{PA}_\text{Q}\propto \omega^{kT/E_\text{ch}}\label{LF-exp}.
\end{align}
Note that Equation~\ref{LF-exp}, in analogy with Equation~\ref{LF-RO}, fails when $kT \approx E_\text{ch}$, i.e. when the carriers can be considered free.  \\

\begin{figure}[h]
 \includegraphics[width=8.5cm]{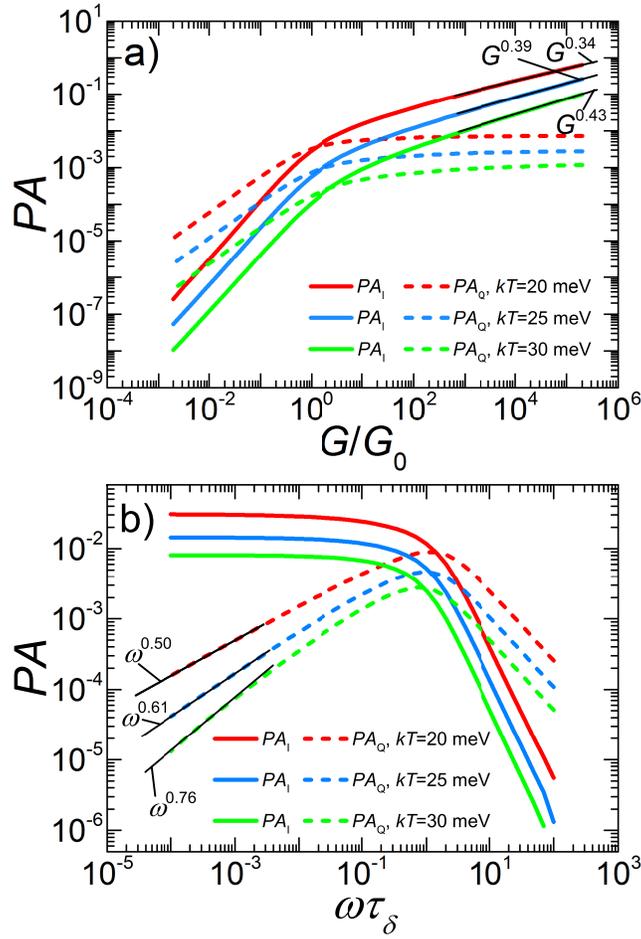}
 \caption{\label{Exp}In-phase ($\textit{PA}_\text{I}$) and quadrature ($\textit{PA}_\text{Q}$) as function of generation $G$ (a) and frequency $\omega$ (b). Simulated results plotted for three temperatures with an exponential trap distribution with $E_\text{ch}=40$ meV. 
}
 \end{figure}

Results from simulations for an exponential distribution with the characteristic energy $E_\text{ch}=40$ meV are illustrated in Figure~\ref{Exp} for three different temperatures. As in Figure~\ref{RO}, the axes are normalized with $\tau_\delta$ and $G_0$, calculated with Equation~\ref{taudelta} and \ref{G0} using Equation \ref{RO-exp} and \ref{beta-exp}. The high intensity in-phase in Figure~\ref{Exp}(a) demonstrates good agreement with Equation~\ref{HI-exp}, with the analytical approximation predicting the slopes {$\gamma_\text{HI}=0.33$} ($kT=20$ meV), 0.38 ($25$ meV), and 0.43 ($20$ meV) compared to the simulated 0.34, 0.39, and 0.43. Likewise, the low frequency quadrature slopes in Figure~\ref{Exp}(b) confirms the usefulness of Equation~\ref{LF-exp}, with the equation predicting {$\gamma_\text{LF}=0.50$} ($20$ meV), 0.63 ($25$ meV), and 0.75 ($20$ meV) in agreement with the simulated 0.50, 0.61, and 0.76. As in the simulations with direct recombination, the results show no indication of dispersion.\\
 
 Simulations (not shown) with different $E_\text{ch}$ for acceptor and donor traps, suggested that the shallower trap distributions dominates the behavior,\cite{gradu} in agreement with Kirchartz { and Nelson}.\cite{kirchartz2012} We also note that applying our model to the decay of charge carriers in an exponential trap distribution after a pulse of photo excitiation, reproduces the (high photoexcitation density) behaviour of $n\propto t^{kT/E_{ch}-1}$ discussed for amorphous semiconductors by Orenstein and Kastner. \cite{orenstein}\\ 

\subsection{Gaussian trap distribution}
\begin{figure}[h]
\includegraphics[width=8.5cm]{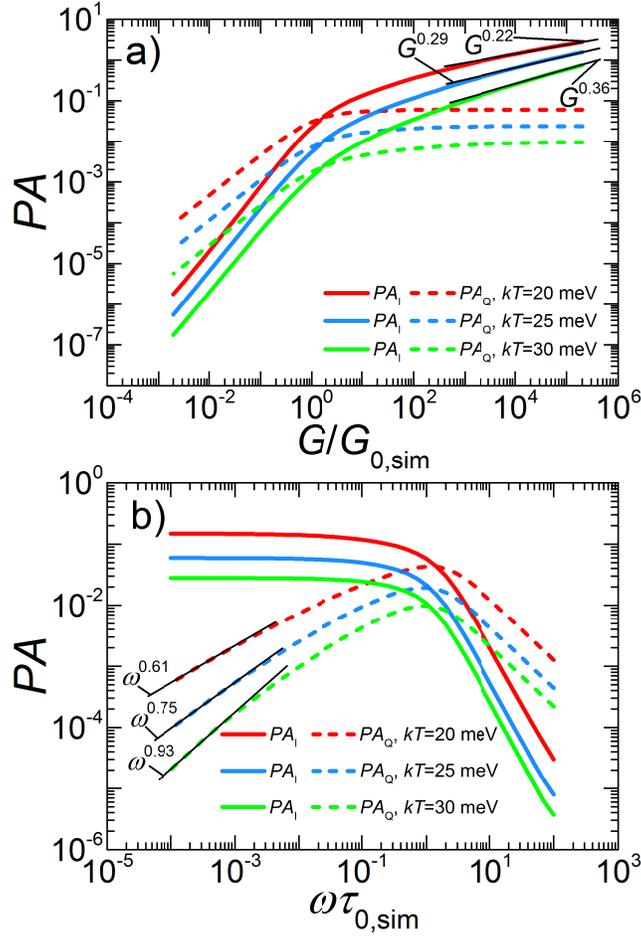}
 \caption{\label{Gauss}In-phase ($\textit{PA}_\text{I}$) and quadrature ($\textit{PA}_\text{Q}$) as function of the generation $G$ (a) and frequency $\omega$ (b). Simulated results plotted for three temperatures with a Gaussian trap distribution, for which $\sigma_\text{G}=0.1$ eV and $E_0=E_\text{C}-40$ meV.}
 \end{figure}

The intensity and frequency dependence for a Gaussian distribution with $E_0=E_\text{C}-\text{40 meV}$ for acceptor traps (for donor traps $E_0=E_\text{V}+40$ meV) and $\sigma_\text{G}=0.1$ eV are depicted in Figure~\ref{Gauss}. 
 As in previous plots the axes are normalized to a unitless scale. However, in the Gaussian case we obtained no analytical approximation for the reaction order. Hence $\tau_\delta$ and $G_0$ could not be calculated as was done for direct recombination and the exponential distribution, instead we used the values $\tau_\text{0,sim}$ and $G_\text{0,sim}$, obtained from the simulated results themselves. The lifetime $\tau_\text{0,sim}$ was defined as the inverse of the frequency where $\textit{PA}_\text{Q}$ has its maximum in the frequency dependence and $G_\text{0,sim}$ as the generation in the intensity dependence where the slope of $\textit{PA}_\text{I}$ is one. \\

As Figure~\ref{Gauss} illustrates, for the Gaussian distribution, in contrast to the exponential distribution, the high intensity in-phase and low frequency quadrature do not exhibit constant slopes. The slopes are described by two functions, {$\gamma_\text{HI}=f(G,T)$} and {$\gamma_\text{LF}=h(\omega,T)$}, for which $\mathrm{d}f/\mathrm{d}G<0$ and $\mathrm{d}h/\mathrm{d}\omega<0$. This corresponds to a reaction order increasing when $\omega$ or $G$ increases. 

\subsection{Agreement of analytical expressions and simulations}
\begin{figure}[h]
 \includegraphics[width=8.5cm]{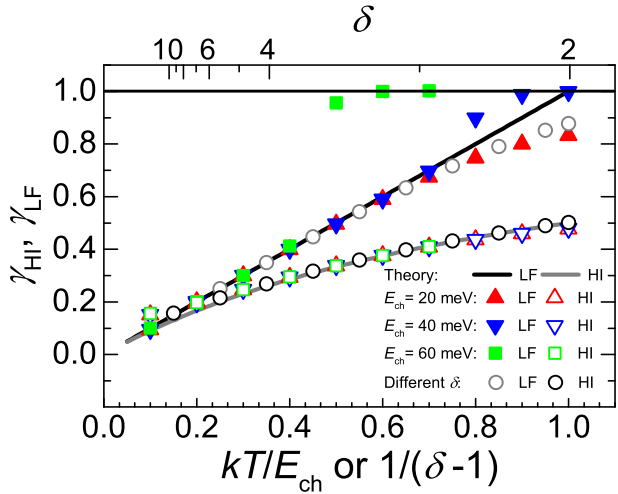}%
 \caption{\label{exp-slope} Analysis of the accuracy of Equation~\ref{HI-RO}, \ref{LF-RO}, \ref{HI-exp} and \ref{LF-exp}. The plots show the value of the high intensity in-phase ({$\gamma_\text{HI}$}) and low frequency quadrature slopes {($\gamma_\text{LF}$)} as obtained from these equations (lines) compared with results from simulations with different reaction orders (circles) and trap-assisted recombination in an exponential distribution with three different $E_\text{ch}$ (triangles and squares). The simulated slopes have been obtained from the in-phase for $2\times10^{4}<G/G_0<2\times10^{5}$ and from the quadrature for $10^{-4}<\omega\tau_\delta<10^{-3}$. Slopes for $T>500$ K not shown due to insufficient accuracy.}
 \end{figure}

Figure~\ref{exp-slope} compares predicted slopes from analytical approximations with simulations. The circles describe simulated results for direct recombination and is thus compared with Equation~\ref{HI-RO} and \ref{LF-RO}. The included recombination orders range from 2 to 7.67. The comparison indicates that Equation~\ref{HI-RO}, describing the high intensity slope, holds reasonably well although a slight deviation starts to occur at very high reaction orders. Equation~\ref{LF-RO} is accurate for high reaction orders but starts to deviate from the simulations when $\delta\approx 2$, as is expected when approaching the logarithmic regime. \\

 The accuracy of Equation~\ref{HI-exp} and \ref{LF-exp}, describing trap-assisted recombination, is compared to simulations for three different $E_\text{ch}$ in Figure~\ref{exp-slope}. The plot demonstrates that the equations are relatively accurate for all trap depths when $kT/E_\text{ch}$ is small. When $kT/E_\text{ch}$ grows two problems appear in the description of the frequency behavior. The two deeper trap distributions, $E_\text{ch}=40$ meV and 60 meV, display a behavior approaching $\textit{PA}_\text{Q}\propto \omega^1$. This arises when the temperature becomes so high that the amount of thermally generated carriers cannot be neglected, disqualifying the use of Equation~\ref{grundRO}. The other problem appears for the shallowest distribution, $E_\text{ch}=20$ meV, which deviates from the predicted line and resembles the logarithmic behavior characteristic of bimolecular recombination. This is expected as the influence of the traps diminish when $kT\approx E_\text{ch}$. 

\section{Conclusions}
We have identified characteristics in measurements of cwPA which enable distinction between trap-assisted recombination in different trap distributions and direct recombination. { The results can be used to obtain information from the slopes (on a log-log scale) of the low frequency quadrature in a frequency dependent measurement and the high intensity in-phase in an intensity dependent measurement. Trap-assisted recombination is revealed by these slopes being temperature dependent. If these slopes are constant throughout the low frequency and high intensity regimes, respectively, an exponential distribution has been found.} Table \ref{identifying} summarizes the results. For an exponential distribution we found relations between the simulated measurement and the mathematical characteristics of the distribution while the functional form of the slopes in the Gaussian case remain unexplored. The formulas for the exponential case were found useful when $kT$ is smaller than $E_\text{ch}$ and not so large that the thermal generation becomes comparable to the photogeneration. 
 Note that the simulations failed to reproduce the impact of dispersion, a goal for future developments. Nonetheless, the results for high intensities and low frequencies are useful and our findings enable study of the trap distributions in a contactless experiment, furthering the study of recombination and energy levels in disordered semiconductors.  \\
\begin{table}[h!]
  \centering
  \caption{Summary of the identifying characteristics of different dominating recombination mechanisms. High intensity dependence ($G\propto$ intensity) for $\textit{PA}_\text{I}$ and low frequency dependence for $\textit{PA}_\text{Q}$. The conditions for high intensity or low frequency are fulfilled when $\omega\tau_\delta\ll1$. See text for discussion of validity range of the expressions.}
    \begin{tabular}{lcc}
    \hline
    \hline
         \parbox[c][3em][c]{4cm}{Recombination type}& ${\textit{PA}_\text{I}}$&${\textit{PA}_\text{Q}}$\\
    \hline
          \parbox[c][4em][c]{3.5cm}{Reaction order $\delta$}& $G^{1/\delta}$ &\parbox[c][4em]{2.2cm}{$\omega^{1/(\delta-1)}$,\\ $\delta>2$} \\ 
         \vspace{-3mm} &&\\
 	\parbox[c][4em][c]{3.5cm}{Exponential trap distribution}& $G^{1/(1+E_\text{ch}/kT)}$  &  \parbox[c][4em][c]{2.2cm}{$\omega^{kT/E_\text{ch}}$,\\ $kT<E_\text{ch}$}\\ 
 	         \vspace{-3mm} &&\\
\parbox[c][4em][c]{3.5cm}{Gaussian trap distribution} &\parbox[c][4em][c]{2.2cm}{$G^{f(G,T)}$, ${\mathrm{d}f}/{\mathrm{d}G}<0$} &\parbox[c][4em][c]{2.2cm}{$\omega^{h(\omega,T)}$,\\${\mathrm{d}h}/{\mathrm{d}\omega}<0$} \\ \hline \hline
    \end{tabular}%
  \label{identifying}%
\end{table}%


\begin{acknowledgments}
Partial financial support from the Academy of Finland through the project No. {279055}, the Magnus Ehrnrooth foundation and the National Graduate School of Nanoscience (NGS-nano) (O.J.S.) is acknowledged. A personal grant from The Waldemar von Frenckell foundation is acknowledged by S.S.
\end{acknowledgments}


\appendix*\section{Calculation of $\textit{PA}_\text{Q}\propto\omega^{1/(\delta-1)}$}
Consider Equation~\ref{grundRO} when $\mathcal{G}(t)=G$ and $\omega\tau_\delta\ll1$, with $\tau_\delta$ defined in equation~\ref{taudelta}. The equation will lead to the stabilization of the density at $N_\text{ss}=(G/b)^{1/\delta}$. If $G$ is large or $\omega$ is small the stabilization will be almost instantaneous. As a measure of this consider the case of the density rising from 0 to $N_\text{ss}$ under the influence of the generation $G$ without recombination. The rise time $t_\text{r}$  can be calculated from 
\begin{equation}
\int_0^{t_\text{r}}G\text{d}t=N_\text{ss} \implies t_\text{r}=\frac{N_\text{ss}}{G}=\frac{G^{1/\delta-1}}{b^{1/\delta}}=\tau_\delta
\end{equation} 
If the rise is assumed instantaneous when $t_\text{r}\ll1/\omega$ the constraint can be written as $\omega\tau_\delta\ll1$. When $\mathcal{G}(t)=0$ the density will decay according to
\begin{align}
&\frac{{\rm d}N(t)}{{\rm d}t}=-b N(t)^\delta\nonumber\\
&\implies N(t)=[b(t+K)(\delta-1)]^{1/(1-\delta)}, 
\label{ndecay}
\end{align}
where $K$ is a constant. The time scale is chosen so that $\mathcal{G}(t)=G$ when $\pi/\omega<t<0$, resulting in the density 

\begin{equation}\label{N-approx}
N(t)=\begin{cases}
(G/b)^{1/\delta},&\frac{\pi}{\omega}<t<0\\
[b(t+K)(\delta-1)]^{1/(1-\delta)},&0<t<\frac{\pi}{\omega}.
\end{cases}
\end{equation}
From demanding continuity at $t=0$ we get $K=G^{1/\delta-1}b^{-1/\delta}(\delta-1)^{-1}=\tau_\delta/(\delta-1)$. To calculate $\textit{PA}_\text{Q}$ Equation~\ref{N-approx} is inserted into Equation~\ref{PAQ-def}, yielding

\begin{align}
\begin{split}
-\textit{PA}_\text{Q}\approx &
\underbrace{\frac{\omega\sigma  d}{\pi}\int_{-\pi/\omega}^{0}\cos{(\omega t)}N(t){\rm d}t}_{\equiv I_1}\\
&+\underbrace{\frac{\omega \sigma d}{\pi}\int_{0}^{\pi/\omega}\cos{(\omega t)}N(t){\rm d}t}_{\equiv I_2}.
\end{split}
\end{align}

The integral $I_1$ can easily be verified to give $I_1=0$ while the second integral can be calculated by a change of variables according to

\begin{align}
I_2&=\frac{\sigma d}{\pi}\int_{0}^{\pi}\cos{(v)}[b(v/\omega+\tau_\delta(\delta-1)^{-1})(\delta-1)]^{1/(1-\delta)}{\rm d}v\nonumber\\
&={\frac{\sigma d}{\pi}\left(\frac{b}{\omega}\right)^{1/(1-\delta)}}{\int_{0}^{\pi}\frac{\cos{(v)}{\rm d}v}{\left(v(\delta-1)+\omega \tau_\delta\right)^{1/(\delta-1)}}}.
\label{intA}
\end{align}

When $\omega\tau_\delta\rightarrow 0$ the integral will be given by 

\begin{equation}
I_2\rightarrow \frac{\sigma d}{\pi}\left(\frac{b(\delta-1)}{\omega}\right)^{1/(1-\delta)}{\int_{0}^{\pi}\cos{(v)}v^{1/(1-\delta)}}{\rm d}v
\end{equation}

The integral ${\int_{0}^{\pi}\cos{(v)}v^{1/(1-\delta)}}{\rm d}v$ converges if $\delta>2$. Thus $\textit{PA}_\text{Q}$ will,  when $\omega\tau_\delta<<1$, depend on the frequency as

\begin{equation}
\textit{PA}_\text{Q}\propto \omega^{1/(\delta-1)}, \text{ when }\delta>2 .
\end{equation}


\end{document}